\documentclass[aps,prb,twocolumn,floatfix,showpacs]{revtex4}
\usepackage{amsmath,amssymb,graphicx,bm}
\begin{document}
\title{Vertex corrections to the mean-field electrical conductivity in strongly disordered electron systems}

 \author{ V. Pokorn\'y} \author{V.  Jani\v{s}}

\affiliation{Institute of Physics, Academy of Sciences of the Czech
  Republic, Na Slovance 2, CZ-18221 Praha 8, Czech Republic}
\email{pokornyv@fzu.cz, janis@fzu.cz}

\date{\today}

%\maketitle

\begin{abstract}
Mean-field theory of non-interacting disordered electron systems is widely and successfully used to describe equilibrium properties  of alloys in the whole range of disorder strengths. It, however, fails to take into account effects of quantum coherence and localizing back-scattering effects when applied to transport phenomena. We present an approximate scheme extending the mean-field theory for one-electron properties in that it offers a formula for the two-particle vertex and the electrical conductivity non-perturbatively  including the leading-order vertex corrections in a way that the approximation remains consistent and the conductivity non-negative in all disorder regimes.    
\end{abstract}

\maketitle %newpage

\section{Introduction}
\label{sec:Intro}

Fluctuations of the spatial distribution of the atomic potential give origin to the zero-temperature resistivity of crystalline solids.  In order to study in detail the impact of fluctuations of the atomic potential on charge transport it is appropriate to neglect all less important agents influencing the response of the electron gas on external electric perturbations and to investigate only scatterings of free electrons on random impurities. It is usual to study the effects of impurity scatterings on an Anderson model of disordered electrons\cite{Anderson58} with a site-independent distribution of the atomic potential.  A straightforward way to investigate this model is to use a perturbation (diagrammatic) expansion in the random atomic potential. Nontrivial and physically interesting results can be, however, reached only via non-perturbative approaches. The most reliable non-perturbative method of describing the effects of disorder on one-electron functions is a mean-field, coherent-potential approximation (CPA). \cite{Elliot74} The coherent potential approximation is nowadays considered as the archetype
of mean-field theories of quantum disordered and interacting systems. Its
generalized form\cite{Janis89} offers one possible interpretation of
equations of motion in the dynamical mean-field theory
(DMFT).\cite{Georges96} The CPA has proved reliable  to produce an accurate
equilibrium electronic structure of disordered systems\cite{Gonis92} as
well as  transport properties of random alloys in a wide range of disorder strength.\cite{Weinberger03} It, however, fails to account for inter-site quantum coherence and
backscattering effects. The CPA is essentially unable to go beyond the
semi-classical description of transport properties qualitatively captured
by the Boltzmann equation. This inability is due to the fact that the CPA
does not include vertex corrections to the one-electron (Drude-type) electrical conductivity.\cite{Velicky69}  In the simplest single-band tight-binding model the electrical conductivity at zero temperature in $\alpha$ direction can be represented in the CPA via the averaged one-electron propagator at the Fermi surface
\begin{equation}\label{eq:Conductivity-Drude}
  \sigma^{(0)}_{\alpha\alpha} = \frac{e^2}{\pi
    N}\sum_{\mathbf{k}}\left|v_\alpha(\mathbf{k}) \right|^2 \left|\Im
    G^R(\mathbf{k}) \right|^2\ 
\end{equation}
where $v_{\alpha}(\mathbf{k})= \partial \epsilon(\mathbf{k})/\partial k_{\alpha}$ is the electron group velocity in direction $\alpha$ and superscript $R$ denotes the retarded part of the averaged one-electron propagator. All the contributions to the electrical conductivity beyond $ \sigma^{(0)}$ originate from coherent propagation of an electron and a hole and are contained in vertex corrections. 

Vertex corrections to the Drude conductivity from Eq.~\eqref{eq:Conductivity-Drude} are
important in various situations. In low dimensions ($d\le2$) or for
sufficiently strong disorder they are responsible for vanishing of diffusion called Anderson localization.\cite{Lee85}  Further on, tunnel conductance or transport through
multilayered  strongly disordered alloys and dirty metals are essentially influenced by vertex corrections.\cite{Itoh99,Hale11} To obtain more realistic results for the electronic
transport in dirty metals one has to go beyond the standard mean-field
conductivity and to develop approximations containing spatial quantum
coherence and backscattering contributions.
%
%\begin{widetext}
\begin{figure}
  \includegraphics[scale=0.65]{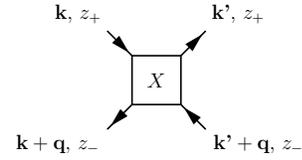}
  \caption{Attribution of complex frequencies $z_{1}, z_{2}$ and 
    momenta $\mathbf{k}, \mathbf{k}', \mathbf{q}$ in the electron-hole diagrammatic representation of the generic two-particle functions $X_{\mathbf{k}\mathbf{k}'}(\mathbf{q}; z_{1},z_{2})$\label{fig:G2-def}}\end{figure}
%\end{widetext}

To calculate the full electrical conductivity from the Kubo formula with the current-current correlation function we need to use the averaged two-particle Green function at the real energy axis.\cite{Janis03a} It can be represented as  
\begin{equation}\label{eq:G2-def}
G^{ab}_{\mathbf{k}\mathbf{k}'}(\mathbf{q},\omega) \equiv G^{ab}_{\mathbf{k}\mathbf{k}'}(\mathbf{q};E + \omega + ia 0^{+}, E + ib 0^{+} ) 
\end{equation}
with a double superscript $ab$ specifying complex half-planes in which the limit to the real axis of the two energy (frequency) variables is taken. We use $a,b= \pm1$ or equivalently $a,b= R,A$ for the upper, lower half-plane for the electron and hole, respectively. Fermionic variables $\mathbf{k},\mathbf{k}'$ are the incoming and the outgoing particle momenta and $\mathbf{q}$ is the difference between the incoming momenta of the particle and the hole in the pair, confer Fig.~\ref{fig:G2-def}. Since the frequencies are not dynamical variables for disordered noninteracting systems we often suppress them in order to simplify notation. Actually, the unspecified frequency (energy) variables are set to the Fermi energy as all calculations are done at zero temperature.  

The contribution  to the electrical conductivity from the free uncorrelated propagation is the Drude term in Eq.~\eqref{eq:Conductivity-Drude}. The contribution from the vertex corrections is expressed via the two-particle vertex defined from the averaged two-electron Green function as  
\begin{multline}\label{eq:G-Gamma}
  G^{ab}_{\mathbf{k}\mathbf{k}'}(\mathbf{q}) = G^a_\mathbf{k} G^b_{\mathbf{k} + \mathbf{q}}\\ \times \left[\delta(\mathbf{k} -\mathbf{k}') + \Gamma^{ab}_{\mathbf{k}\mathbf{k}'}(\mathbf{q})    G^a_{\mathbf{k}'} G^b_{\mathbf{k}' + \mathbf{q}}\right]\ .
\end{multline}
The vertex corrections to the zero-temperature conductivity then read\cite{Mahan90} 
\begin{multline}\label{eq:Conductivity-correction}
  \Delta\sigma_{\alpha\alpha} = \frac{e^2}{2\pi
    N^2}\sum_{\mathbf{k}\mathbf{k}'} v_\alpha(\mathbf{k})
  v_\alpha(\mathbf{k}')\left\{
    \left|G^R_\mathbf{k}\right|^2\Gamma^{AR}_{\mathbf{k}\mathbf{k}'}
    \left| G^R_{\mathbf{k}'}\right|^2  \right. \\ \left.
   - \Re\left[\left(G^R_\mathbf{k}\right)^2
      \Gamma^{RR}_{\mathbf{k}\mathbf{k}'}
      \left(G^R_{\mathbf{k}'}\right)^2 \right] \right\}\ .
  \end{multline}
We suppressed the bosonic momentum variable $q=0$ and frequency $\omega=0$ as we will do throughout the paper, when no confusion may arise,  for variables the values of which are set to zero.    

Troubles with the decoupling of the full conductivity into the Drude term and vertex corrections emerge in the limit of strong disorder and low dimensions where the two-particle corrections with negative sign may overweigh the positive one-electron contribution. There is not a universal guarantee that the total conductivity stays non-negative in the whole range of the disorder strength. Even systematic corrections to the mean-field conductivity derived from the limit to high spatial dimensions may lead in three-dimensional models to negative values of conductivity.\cite{Janis01a,Janis01b} Since the mean-field approximation is consistent for one-electron properties in the whole range of the disorder strength, it is desirable to have an analogous consistency also for transport properties and electrical conductivity with leading-order vertex corrections applicable on realistic random systems in three dimension.  

It is the aim of this paper to present a scheme how to reach an approximate expression for the electrical conductivity in disordered noninteracting systems extending the mean-field expression by involving vertex corrections so that to keep the conductivity non-negative. We use the asymptotic limit to high spatial dimensions to derive an approximate formula for the electrical conductivity with vertex corrections delivering non-negative results. The present paper is an improvement upon the calculation of the vertex corrections to the electrical conductivity derived from the asymptotic limit to high spatial dimensions in Refs.~\onlinecite{Janis01a,Janis01b} in that we keep the full structure of the Brillouin zones of low-dimensional lattices and hence treat momentum convolutions there more realistically.

\section{Electrical conductivity from a symmetric Bethe-Salpeter equation}

To derive a non-perturbative expression for the electrical conductivity we use a representation of the full two-particle vertex via a Bethe-Salpeter equation in the electron-hole channel. Using the allocation of variables of two-particle quantities from Fig.~\ref{fig:G2-def}  and denoting the irreducible vertex in the electron-hole channel $\Lambda$ we have 
\begin{multline}\label{eq:BS-Gamma}
\Gamma^{ab}_{\mathbf{k}\mathbf{k}'}(\mathbf{q}) = \Lambda^{ab}_{\mathbf{k}\mathbf{k}'}(\mathbf{q}) \\ + \frac 1N \sum_{\mathbf{k}''} \Lambda^{ab}_{\mathbf{k}\mathbf{k}''}(\mathbf{q})  G^a_{\mathbf{k}''} G^b_{\mathbf{k}'' + \mathbf{q}} \Gamma^{ab}_{\mathbf{k}''\mathbf{k}'}(\mathbf{q}) \ .
\end{multline}
This Bethe-Salpeter equation can be extended to the full two-particle Green function by using the definition in Eq.~\eqref{eq:G2-def}.  The two-particle Green function can then be determined from the electron-hole irreducible vertex  as
\begin{multline}\label{eq:BS-G2}
G^{ab}_{\mathbf{k}\mathbf{k}'}(\mathbf{q}) = G^a_\mathbf{k} G^b_{\mathbf{k} + \mathbf{q}}\\ \times \left[\delta(\mathbf{k} -\mathbf{k}')  + \frac 1N \sum_{\mathbf{k}''} \Lambda^{ab}_{\mathbf{k}\mathbf{k}''}(\mathbf{q})  G^{ab}_{\mathbf{k}''\mathbf{k}'}(\mathbf{q}) \right] \ .
\end{multline}
This is an integral equation in momentum space with a kernel $\Lambda^{ab}_{\mathbf{k}\mathbf{k}'}(\mathbf{q}) G^a_{\mathbf{k}'} G^b_{\mathbf{k}' + \mathbf{q}}$ where fermionic momenta $\mathbf{k}$ and $\mathbf{k}'$ are the active ones. This integral kernel is asymmetric in their active variables even for the difference momentum $q=0$. We can, however, transform the above Bethe-Salpeter equation to a symmetric one allowing for an easier diagonalization and non-perturbative solution.  To do so we redefine two-particle functions in momentum space by multiplying them from right and left by one-electron propagators 
\begin{equation}\label{eq:Xtilde}
\widehat{X}^{ab}_{\mathbf{k}\mathbf{k}'}(\mathbf{q}) = G^{a}_{\mathbf{k} }  X^{ab}_{\mathbf{k}\mathbf{k}'} (\mathbf{q}) G^b_{\mathbf{k}' + \mathbf{q}}\ .
\end{equation}
The one-electron propagators are now absorbed in the rescaled vertex that forms a symmetric kernel for a Bethe-Salpeter equation for the two-particle vertex
\begin{equation}\label{eq:BS-Gamma-symmetric}
\widehat{\Gamma}^{ab}_{\mathbf{k}\mathbf{k}'}(\mathbf{q}) = \widehat{\Lambda}^{ab}_{\mathbf{k}\mathbf{k}'}(\mathbf{q}) + \frac 1N \sum_{\mathbf{k}''} \widehat{{\Lambda}}^{ab}_{\mathbf{k}\mathbf{k}''}(\mathbf{q})  \widehat{\Gamma}^{ab}_{\mathbf{k}''\mathbf{k}'}(\mathbf{q}) \ .
\end{equation}
 Analogously to this Bethe-Salpeter equation we obtain a symmetrized Bethe-Salpeter equation for the full two-particle Green function the solution of which we formally represent as 
\begin{equation}\label{eq:BS-G2_{symmetric}}
G^{ab}_{\mathbf{k}\mathbf{k}'}(\mathbf{q}) = G^b_{\mathbf{k} + \mathbf{q}}\left[ 1 - \widehat{\Lambda}^{ab}(\mathbf{q}) \star\right]^{-1}_{\mathbf{k}\mathbf{k}'} G^{a}_{\mathbf{k}'}\ ,
\end{equation}
where $\star$ stands for the appropriate matrix multiplication in the $eh$ scattering channel used in Eq.~\eqref{eq:BS-G2}. 

This representation of the two-particle Green function can be used to derive a non-perturbative representation of the electrical conductivity that is free of the decomposition into the Drude term and vertex corrections. We obtain for the static optical conductivity at zero temperature 
\begin{multline}\label{eq:sigma-nonperturbative}
\sigma_{\alpha\beta} = \frac{e^{2}}{2\pi N^{2}}\sum_{\mathbf{k}\mathbf{k}'} v_{\alpha}(\mathbf{k})\left\{ G^{A}_{\mathbf{k}}\left[1 - \widehat{\Lambda}^{RA} \star\right]^{-1}_{\mathbf{k}\mathbf{k}'} G^{R}_{\mathbf{k}'} \right.\\ \left.
- \Re \left( G^{R}_{\mathbf{k}}\left[1 - \widehat{\Lambda}^{RR}\star \right]^{-1}_{\mathbf{k}\mathbf{k}'} G^{R}_{\mathbf{k}'}\right)\right\} v_{\beta}(\mathbf{k}'
)\ .\end{multline}
The expression for the electrical conductivity via the electron-hole irreducible vertex $\widehat{\Lambda}$ in Eq.~\eqref{eq:sigma-nonperturbative} leads to non-negative results if  the inverse operator on its  right-hand side is correctly calculated or approximated. Its perturbation expansion, or separating the even part of the inverse operator, leads to  decomposition of the conductivity into the Drude term (even contribution) and the vertex corrections (odd contribution).  It is hence mandatory to use a non-perturbative evaluation or approximation for the inverse operator in Eq.~\eqref{eq:sigma-nonperturbative} to guarantee non-negative results for all strengths of disorder.

\section{Expansion around mean field}
\label{sec:CPA-beyond}

The representation from Eq.~\eqref{eq:sigma-nonperturbative} offers a way how to calculate the electrical conductivity non-perturbatively without the necessity to decompose it into the mean-field one-electron contribution and the two-particle vertex corrections. To guarantee non-negative results for the conductivity one has to resolve the inverse in momentum space non-perturbatively. Ideal would be to diagonalize the irreducible vertex exactly. It is, however, possible only numerically in a discretized space with only a few points in the Brillouin zone. In realistic situations one need not evaluate the inverse operator exactly, since the vertex corrections are in most situations small. So that to keep the effort in calculating the electrical conductivity with vertex corrections within reasonable limits, we resort to a non-perturbative approximation of the matrix inversion in Eq.~\eqref{eq:sigma-nonperturbative}. We use the asymptotic limit to high spatial dimensions as a guide in the selection of the appropriate manageable approximate scheme. The vertex corrections can be systematically controlled in this way.

\subsection{Leading-order non-local electron-hole vertex}
\label{sec:Leading_order}

To determine the leading-order vertex correction to the electrical conductivity we use an expansion around the mean-field solution, being exact in $d=\infty$. The mean-field solution becomes exact when the off-diagonal part of the full one-electron Green function vanishes. It leads to a local self-energy $\Sigma(z)$. A small parameter for the expansion around the mean-field solution is the off-diagonal mean-field Green function that can be represented in momentum space as      
\begin{equation}\label{eq:G-off}
  \overline{G}(\mathbf{k},\zeta) = \frac 1{\zeta - \epsilon(\mathbf{k})} - \int
  \frac{d\epsilon \rho(\epsilon)}{\zeta -\epsilon}
\end{equation}
with $\zeta = z - \Sigma(z)$. The mean-field approximation is consistent only when the one-electron Green function is everywhere replaced by its local element. It applies in particular on two-particle quantities. The only consistent and unambiguous two-particle vertex within the mean-field approximation is the local one 
\begin{equation}\label{eq:gamma-CPA}
\gamma^{ab} = \frac{\lambda^{ab}}{1 - \lambda^{ab}G^{a}G^{b}}
\end{equation}
where we used the two-particle irreducible vertex $\lambda$. It is related to the one-electron self-energy via a Ward identity.\cite{Velicky69,Vollhardt80a} The Ward identity is channel-dependent.  We have in the electron-hole and electron-electron channels\cite{Janis10} 
\begin{subequations}\label{eq:WI}
  \begin{align}
    \lambda^{RA} &= \frac{\Im\Sigma^{R}}{\Im G^{R}} = \frac 1
    {{\chi}^{RA}(\mathbf{0})}\ , \\ \lambda^{RR} &=
    \frac{\Sigma^{R\prime}}{ G^{R\prime}} = \frac {Z^{R}}
    {{\chi}^{RR}(\mathbf{0})}\   
\end{align}\end{subequations}%
where we denoted $\Sigma^{R\prime} = \partial\Sigma^{R}(\omega)/\partial\omega|_{\omega=0}$,
$G^{R\prime} = \partial G^{R}(\omega)/\partial\omega|_{\omega=0}$, and $Z^{R}=\Sigma^{R\prime} /(\Sigma^{R\prime} - 1) $.
We also related the irreducible vertices with the homogeneous part of the non-local two-particle bubble, a convolution of two one-electron propagators  
\begin{equation}\label{eq:chi-CPA}
\chi(\zeta,\zeta';\mathbf{q}) = N^{-1}\sum_{\mathbf{k}}
  G(\mathbf{k},\zeta) G(\mathbf{k} + \mathbf{q},\zeta')\ .
\end{equation}
This bubble is non-local even in the mean-field limit ($d=\infty$). 

When trying to extend the mean-field approximation also to the non-local two-particle functions, the results are no longer consistent, since the non-local vertex derived with the aid of the Ward identity, used in the CPA, does not cover all the non-vanishing contributions to two-particle functions in $d=\infty$. This ambiguity is a consequence of the fact that two (linear) operations do not commute, namely (functional) derivative of Green functions with respect to external sources and the limit $d\to\infty$.\cite{Janis99a}  If we introduce the off-diagonal (non-local) part of the two-particle bubble
\begin{align}\label{eq:chi_bar-CPA}
\overline{\chi}^{ab}(\mathbf{q}) &= N^{-1}\sum_{\mathbf{k}}
  \overline{G}^{a}(\mathbf{k}) \overline{G}^{b}(\mathbf{k} + \mathbf{q}) 
 \nonumber \\  
  & = \chi^{ab}(\mathbf{q}) - G^{a}G^{b} 
\end{align}
we can represent the leading non-local vertex correction to the local mean-field vertex in a form\cite{Janis01a,Janis01b} 
\begin{multline}\label{eq:vertex-high}
  \Gamma_{\mathbf{k}\mathbf{k}'}(\zeta_+,\zeta_-;\mathbf{q}) \doteq
  \gamma(\zeta_+,\zeta_-) + \lambda(\zeta_+,\zeta_-)\\ \times
  \left[\frac{\gamma(\zeta_+,\zeta_-) \bar{\chi}(\zeta_+,\zeta_-;\mathbf{q})} {1 -
      \lambda(\zeta_+,\zeta_-) \chi(\zeta_+,\zeta_-;\mathbf{q})}\right. \\ \left. +\
    \frac{\gamma(\zeta_+,\zeta_-)\bar{\chi}(\zeta_+,\zeta_-;\mathbf{q} + \mathbf{k} +    \mathbf{k}')} {1 -
      \lambda(\zeta_+,\zeta_-) \chi(\zeta_+,\zeta_-;\mathbf{q} + \mathbf{k} +    \mathbf{k}')}\right]\ .
\end{multline}
 Standardly the full momentum-dependent mean-field vertex contains only the first two terms on the right-hand side of Eq.~\eqref{eq:vertex-high} independent of the fermionic momenta $\mathbf{k}$ and $\mathbf{k}'$. Due to symmetry reasons they  do not contribute to the electrical conductivity.  It is the third term that generates the leading-order vertex corrections to the electrical conductivity and is responsible for the so-called weak localization.\cite{Vollhardt80b} 

To avoid addition of the vertex corrections to the mean-field conductivity we use the Bethe-Salpeter equation \eqref{eq:BS-G2_{symmetric}} and express the conductivity via the electron-hole irreducible vertex $\widehat{\Lambda}$.  Since we expand around the mean-field solution, we must use a modified Bethe-Salpeter equation suppressing multiple scatterings on the same site so that to avoid double counting of scattering effects. We denote the modified irreducible vertex  $\overline{\Lambda}$.  A modified Bethe-Salpeter equation for the full vertex  reads\cite{Janis05a}
\begin{multline}\label{eq:BS-eh}
  \Gamma_{\mathbf{k}\mathbf{k}'}(z_+,z_-;\mathbf{q}) =
  \overline{\Lambda}_{\mathbf{k}\mathbf{k}'}(z_+,z_-;\mathbf{q})\\ +
  \frac 1N\sum_{\mathbf{k}''}
  \overline{\Lambda}_{\mathbf{k}\mathbf{k}''}(z_+,z_-;\mathbf{q})
  \overline{G}_+(\mathbf{k}'') \overline{G}_-(\mathbf{k}'' + \mathbf{q})\\
  \times\Gamma_{\mathbf{k}''\mathbf{k}'}(z_+,z_-;\mathbf{q}) \ .
\end{multline}
The leading order contribution to the off-diagonal part of the irreducible vertex in high spatial dimensions is just the last term on the right-hand side of Eq.~\eqref{eq:vertex-high}. We then have
\begin{equation}\label{eq:Lambda-bar-eh}
\overline{\Lambda}^{ab}_{\mathbf{k}\mathbf{k}'}(\mathbf{q}) = \gamma^{ab} \left[1 + \frac{\gamma^{ab} \overline{\chi}^{ab}(\mathbf{k} + \mathbf{k}'+ \mathbf{q})}{1 - \gamma^{ab} \overline{\chi}^{ab}(\mathbf{k} + \mathbf{k}'+ \mathbf{q})} \right]\ .
\end{equation}
To calculate the conductivity with the aid of vertex $\overline{\Lambda}$ and a symmetrized Bethe-Salpeter equation we redefine the two-electron Green function 
\begin{multline}\label{eq:BS-G2-bar}
 \overline{G}^{ab}_{\mathbf{k}\mathbf{k}'}(\mathbf{q}) = G^b_{\mathbf{k} + \mathbf{q}}\\ \times \left[\delta(\mathbf{k} -\mathbf{k}') + \overline{G}^a_{\mathbf{k}} \Gamma^{ab}_{\mathbf{k}\mathbf{k}'}(\mathbf{q})  \overline{G}^b_{\mathbf{k}' + \mathbf{q}}   \right] G^a_{\mathbf{k}'} 
\end{multline}
that does not change the long-range diffusive behavior of the full two-particle Green function. This constrained function can be represented as a solution of a symmetrized Bethe-Salpeter equation
\begin{equation}\label{eq:BS-G2-bar_{symmetric}}
\overline{G}^{ab}_{\mathbf{k}\mathbf{k}'}(\mathbf{q}) = G^b_{\mathbf{k} + \mathbf{q}}\left[ 1 - \widehat{\overline{\Lambda}}\phantom{}^{ab}(\mathbf{q}) \star\right]^{-1}_{\mathbf{k}\mathbf{k}'} G^{a}_{\mathbf{k}'}
\end{equation}
with a kernel  that we denote for further calculations as $K^{-1} = 1 - \widehat{\overline{\Lambda}}$. %

 We use the constrained two-electron Green function $\overline{G}^{ab}$ in the calculation of the electrical conductivity. The result is not changed by this substitution. From the symmetrized Bethe-Salpeter equation~\eqref{eq:BS-G2-bar}  we obtain
\begin{multline}\label{eq:sigma-bar-nonperturbative}
\sigma_{\alpha\beta} = \frac{e^{2}}{2\pi N^{2}}\sum_{\mathbf{k}\mathbf{k}'} v_{\alpha}(\mathbf{k})\left\{ G^{A}_{\mathbf{k}}\left[1 - \widehat{\overline{\Lambda}}\phantom{}^{RA} \star\right]^{-1}_{\mathbf{k}\mathbf{k}'} G^{R}_{\mathbf{k}'} \right.\\ \left.
- \Re \left( G^{R}_{\mathbf{k}}\left[1 - \widehat{\overline{\Lambda}}\phantom{}^{RR}\star \right]^{-1}_{\mathbf{k}\mathbf{k}'} G^{R}_{\mathbf{k}'}\right)\right\} v_{\beta}(\mathbf{k}'
)\ .\end{multline}
The off-diagonal propagator $\overline{G}$ is the  fundamental parameter in the expansion around the mean-field limit. It makes this expansion consistent and moreover
it makes the calculation of corrections to the mean-field result numerically more stable.  It is preferable to use the full local mean-field vertex $\gamma^{ab}$ instead the irreducible one $\lambda^{ab}$ in all formulas of the expansion around mean field, since  the latter contains a pole in the $RR$ ($AA$) channel that is cancelled in the former one. Note that the leading-order vertex corrections calculated from an expansion of the right-hand side of Eq.~\eqref{eq:sigma-bar-nonperturbative} coincide with the leading corrections to the mean-field conductivity derived  in Ref.~\onlinecite{Janis10}.   
 
The explicit representation for the integral kernel from Eq.~\eqref{eq:sigma-bar-nonperturbative} to be inversed is  
\begin{multline}\label{eq:A-inverse}
\left(K^{ab}\right)^{-1}_{\mathbf{k}\mathbf{k}'}(\mathbf{q}) \equiv  \left(L^{ab}\right)^{-1}_{\mathbf{k}\mathbf{k}'} (\mathbf{q})  - \gamma^{ab} \overline{G}^{a}(\mathbf{k}) \overline{G}^{b}(\mathbf{k}' + \mathbf{q}) \\
= \delta_{\mathbf{k},\mathbf{k}'} -  \gamma^{ab} \overline{G}^{a}(\mathbf{k}) \overline{G}^{b}(\mathbf{k}' + \mathbf{q})S^{ab}(\mathbf{k} + \mathbf{k}'+ \mathbf{q}) \\ -  \gamma^{ab} \overline{G}^{a}(\mathbf{k}) \overline{G}^{b}(\mathbf{k}' + \mathbf{q})
\end{multline}
with $\mathbf{k}$ and $\mathbf{k}'$ as active variables. We denoted 
\begin{equation}\label{eq:S_function-def}
S^{ab}(\mathbf{Q}) =\gamma^{ab}\overline{\chi}^{ab}(\mathbf{Q})/\left[1 - \gamma^{ab} \overline{\chi}^{ab}(\mathbf{Q})\right]\ .
\end{equation} 

One can try to invert numerically exactly the matrix from Eq.~\eqref{eq:A-inverse}, but it is a rather demanding task. Since we expect that the mean-field conductivity dominates within the energy bands with relevant vertex corrections only close to the band edges we can resort to an approximate matrix inversion so that non-negativity of the result is preserved and the bulk conductivity not much affected. 

\subsection{Approximate solution}
\label{sec:Approx}

The dominant contribution to the irreduciblle vertex $\overline{\Lambda}$ from Eq.~\eqref{eq:Lambda-bar-eh} comes from the pole  in  
 $S^{RA}(\mathbf{Q})$ for $Q=0$. It means that we can single out fermionic momenta $\mathbf{k}, \mathbf{k}'$ in the inversion so that   $\mathbf{k} + \mathbf{k}'+ \mathbf{q} \approx 0$. Then the pre-factor $\overline{G}_{\mathbf{k}} \overline{G}_{\mathbf{k}' + \mathbf{q}}$ at function $S^{RA}(\mathbf{k} + \mathbf{k}'+ \mathbf{q})$ in Eq.~\eqref{eq:A-inverse} becomes relevant only in the vicinity of the pole, that is for $\mathbf{k} + \mathbf{k}'+ \mathbf{q} \approx 0$.  We further get rid of the dependence of the pre-factor on the fermionic variable in that we replace it by its average over the Brillouin zone, namely $N^{-1}\sum_{\mathbf{k}}\overline{G}_{\mathbf{k}}\overline{G}_{-\mathbf{k}} $. This simplification leads to the inverse of a reduced matrix 
\begin{equation}\label{eq:L_inverse-approximate}
L^{-1}_{\mathbf{k}\mathbf{k}'} (\mathbf{q}) \doteq  \delta_{\mathbf{k}\mathbf{k}'} - \gamma \overline{\chi}(\mathbf{0}) S(\mathbf{k} + \mathbf{k}'+ \mathbf{q})\ .
\end{equation}
Although this approximation is justified for vertex $\overline{\Lambda}^{AR}$, since only the electron-hole channel contains the pole,  we use it, due to consistency, also for vertex $\overline{\Lambda}^{RR}$. 

Inverse of the matrix on the right-hand side of Eq.~\eqref{eq:L_inverse-approximate} can be explicitly evaluated by using a Fourier transformation diagonalizing convolutions (correlations) resulting from multiplication in the matrix inversion in momentum space. We introduce  a Fourier transform in a $d$-dimensional momentum space as follows
\begin{align*}%\label{eq:Fourier-def}
\widetilde{f}(\mathbf{x}) &= \int d\mathbf{q}\ e^{i\mathbf{q}\cdot\mathbf{x}}f(\mathbf{q})\ , \\  f(\mathbf{q}) &= \int\frac{d\mathbf{x}}{(2\pi)^{d}}\  e^{- i \mathbf{q}\cdot\mathbf{x}}  \widetilde{f}(\mathbf{x}) \ .
\end{align*} 
We can now explicitly represent the inverse matrix
 \begin{multline}\label{eq:L-approximate}
L^{ab}_{\mathbf{k}\mathbf{k}'} (\mathbf{q})  = \int \frac{d \mathbf{x} }{(2 \pi)^{d}}\\ 
\left[ \frac{ e^{-i(\mathbf{k} - \mathbf{k}')\cdot\mathbf{x}} + \gamma^{ab} \overline{\chi}^{ab} \widetilde{S}^{ab}(\mathbf{-x})e^{-i( \mathbf{k} + \mathbf{k}' + \mathbf{q})\cdot\mathbf{x}}}{1 -  \left(\gamma^{ab} \overline{\chi}^{ab}\right)^{2} \widetilde{S}^{ab}(\mathbf{x})\widetilde{S}^{ab}(\mathbf{-x})} \right]\ ,
 \end{multline}
%
%\begin{multline}\label{eq:Lambda_eh-solution}
%^{(eh)}\Lambda^{ab}_{\mathbf{k}\mathbf{k}'}(\mathbf{q}) = \lambda^{ab}
%\int\frac{d\mathbf{x}}{(2\pi)^{d}}\ \widetilde{S}(\mathbf{x})\\ \times
%\frac{e^{-i(\mathbf{k} + \mathbf{k}'+ \mathbf{q})\cdot \mathbf{x}} + \lambda^{ab} G^{a}G^{b}\widetilde{S}(\mathbf{x})e^{-i(\mathbf{k} - \mathbf{k}')\cdot \mathbf{x}}  }{\left(1 - \lambda^{ab}G^{a}G^{b}\widetilde{S}(\mathbf{x})\right)\left(1 + \lambda^{ab}G^{a}G^{b}\widetilde{S}(\mathbf{x})\right)}
%\end{multline}
as well as 
\begin{multline}\label{eq:A-solution}
K^{ab}_{\mathbf{k}\mathbf{k}'} = L^{ab}_{\mathbf{k}\mathbf{k}'} (\mathbf{q})  + \frac{\gamma^{ab}}{1 - \gamma^{ab}\left\langle \overline{G}^{b}L^{ab} \overline{G}^{a}\right\rangle(\mathbf{q})} \\  \times 
\int \frac{d\mathbf{x}}{(2\pi)^{d}}   \widetilde{\overline{G}}^{a}(\mathbf{x}) \left[ \frac{e^{-i\mathbf{k}\cdot\mathbf{x}} + \gamma^{ab}\overline{\chi}^{ab}\widetilde{S}^{ab}(\mathbf{x})e^{-i(\mathbf{k} + \mathbf{q})\cdot\mathbf{x}}}{1 - \left(\gamma^{ab}\overline{\chi}^{ab}\widetilde{S}^{ab}(\mathbf{x})\right)^{2}}\right]\\ \times
\int \frac{d\mathbf{y}}{(2\pi)^{d}} \left[ \frac{ e^{i(\mathbf{k}'+\mathbf{q})\cdot\mathbf{y}} + \gamma^{ab}\overline{\chi}^{ab}\widetilde{S}^{ab}(\mathbf{y})e^{-i\mathbf{k}'\cdot\mathbf{y}}}{1 - \left(\gamma^{ab}\overline{\chi}^{ab}\widetilde{S}^{ab}(\mathbf{y})\right)^{2}}\right] \widetilde{\overline{G}}^{b}(\mathbf{y}) \ ,
\end{multline}
where we denoted
\begin{multline}\label{eq:GLG-averaged}
\left\langle \overline{G}^{b}L^{ab} \overline{G}^{a}\right\rangle( \mathbf{q})  = \frac 1{N^{2}}\sum_{\mathbf{k}\mathbf{k}'} \overline{G}^{b}_{\mathbf{k} +\mathbf{q}} L^{ab}_{\mathbf{k}\mathbf{k}'} (\mathbf{q})  \overline{G}^{a}_{\mathbf{k}'}  \\
 = \int\frac{d\mathbf{x}}{(2\pi)^{d}}  \widetilde{\overline{G}}^{a}(\mathbf{x}) \widetilde{\overline{G}}^{b}(\mathbf{x}) \left[ \frac{ e^{i\mathbf{q}\cdot\mathbf{x}} + \gamma^{ab}\overline{\chi}^{ab}\widetilde{S}^{ab}(\mathbf{x})}{1 - \left(\gamma^{ab }\overline{\chi}^{ab }\widetilde{S}^{ab}(\mathbf{x})\right)^{2}}\right] \ .
\end{multline}
We suppressed the frequency variables, the (infinitesimal) imaginary parts of which are indicated by the superscripts. 

We further introduce Fourier transforms of velocities as 
$\widetilde{vG}^{a}(\mathbf{x}) = \int d\mathbf{k} v(\mathbf{k}) G^{a}_{\mathbf{k}}e^{i\mathbf{k}\cdot\mathbf{x}}$ so that to reach a representation of the electrical conductivity containing the leading-order vertex corrections to the mean-field Drude conductivity in a form guaranteeing its non-negativity 
\begin{multline}\label{eq:sigma-final}
\sigma = \frac{e^{2}}{2\pi}\int \frac{d \mathbf{x}}{(2\pi)^{d}} \left[\frac{\widetilde{vG}^{A}(\mathbf{x})\widetilde{vG}^{R}(\mathbf{-x})}{1 + \gamma^{RA}\overline{\chi}^{RA} \widetilde{S}^{RA}(\mathbf{x})} \right. \\ \left. - \Re\left(\frac{\widetilde{vG}^{R}(\mathbf{x})\widetilde{vG}^{R}(\mathbf{-x})}{1 + \gamma^{RR}\chi^{RR} \widetilde{S}^{RR}(\mathbf{x})} \right)\right] \equiv \sigma^{RA} - \Re\sigma^{RR}\\ 
= \frac{e^{2}}{2\pi} \left\{\left\langle v  G^{A} L^{RA} v G^{R}\right\rangle - \Re \left\langle v G^{R} L^{RR} v G^{R}\right\rangle
\right\}\ .\end{multline}
Note that $\widetilde{vG}(\mathbf{x}) = -\widetilde{vG}(-\mathbf{x})$ while $\widetilde{S}(\mathbf{x}) = \widetilde{S}(\mathbf{-x})$. It is the first term on the right-hand side of Eq.~\eqref{eq:sigma-final} that is dominant within the band with non-zero imaginary part of the self-energy and $\sigma^{RA} \ge |\sigma^{RR}| \ge 0$.

The above approximation is consistent if $\left\langle \overline{G}^{A}L^{RA} \overline{G}^{R}\right\rangle(\mathbf{q}) $ remains positive for $q = 0$. It is fulfilled if a stability condition 
\begin{equation}\label{eq:Conductivity-Stability}
2\ge   \int\frac{d\mathbf{q}}{1 - \gamma^{RA}\overline{\chi}^{RA}(\mathbf{q})}
\end{equation}
is satisfied. This poses a restriction on the disorder strength measured by the full local mean-field vertex $\gamma$ for which this approximation is consistent and realiable.

\section{Electron-hole correlation function and diffusion pole}
\label{sec:Diffusion}

The proposed approximate scheme of inverting the kernel in the Bethe-Salpeter equation so that the electrical conductivity never goes negative can also be used to determine the electron-hole correlation function. Since we are interested only in the diffusion pole of this function we redefine it here so that to keep the polar structure but to make the explicit expression in the expansion around the mean-field limit as simple as possible. We hence use the off-diagonal one-electron propagators to define a constrained electron-hole correlation function 
\begin{equation} \label{eq:DRF-Phi} 
\overline{\Phi}^{RA}_E(\mathbf{q},\omega) =
\frac 1{N^{2}}\sum_{\mathbf{k}\mathbf{k}'}
\overline{G}^A_{\mathbf{k} + \mathbf{q}}\left[ 1 - \widehat{\overline{\Lambda}}\phantom{}^{RA}(\mathbf{q}) \star\right]^{-1}_{\mathbf{k}\mathbf{k}'} \overline{G}^{R}_{\mathbf{k}'}\ .
\end{equation}
As is well known this function contains the so-called diffusion pole when the Ward identity between the self-energy and the irreducible vertex is obeyed for all frequencies.\cite{Janis03a}  Its mean-field, local version is expressed in Eq.~\eqref{eq:WI}. The full, frequency-dependent Ward identity holds only in the mean-field limit. Once we go beyond and introduce non-local corrections to the irreducible vertex, the full Ward identity is in conflict with analyticity of the self-energy and integrability of the diffusion pole.\cite{Janis04a,Janis09} When we approximate the non-local part of the irreducible vertex  from which we then determine the self-energy, the Ward identity can be obeyed only in the static limit, that is, for zero difference frequency between the electron and the hole.  The exact low-energy asymptotics of the electron-hole correlation function with the static Ward identity reads\cite{Janis05b}   
\begin{equation}\label{eq:Phi-high-dim}
\overline{\Phi}^{RA}_E(\mathbf{q},\omega)\approx  \frac {\overline{\Phi}_{0}} {-i A_E\omega
+ D_E^0(\omega)\mathbf{q}^2}
\end{equation}
where $A_{E}\ge 1$ is a constant directly proportional to the disorder strength. This electron-hole correlation function contains the diffusion pole of order $\omega^{-1}$ for $q=0$, but its weight decreases with increasing disorder strength and vanishes in the localized phase.\cite{Janis05b}  Physical consistency dictates that this low-energy behavior of the electron-hole correlation function should be reproduced also in the proposed approximation. The electron-hole correlation function in this approximation has an explicit representation 
\begin{equation}\label{eq:EH-CF}
\overline{\Phi}^{RA}_E(\mathbf{q}) = \frac{\left\langle \overline{G}^{A} L^{RA} \overline{G}^{R}\right\rangle(\mathbf{q}) }{1 - \gamma^{RA}\left\langle \overline{G}^{A} L^{RA} \overline{G}^{R}\right\rangle(\mathbf{q}) }\ . 
\end{equation}

In an approximation with a non-local irreducible vertex  we must go beyond the local mean-field self-energy to guarantee the low-energy asymptotics of Eq.~\eqref{eq:Phi-high-dim}. To make the proposed approximation consistent with the diffusion pole  we have to correct the one-electron self-energy appropriately. We can use the static  (zero-frequency) Ward identity, that complies with this type of approximation,  to determine the imaginary part of the self-energy from the irreducible vertex  
\begin{equation}\label{eq:WI-Im}
\Im \Sigma^{R}(\mathbf{k}) = \frac 1{N} \sum_{\mathbf{k}'} \Lambda^{RA}_{\mathbf{k}\mathbf{k}'}\Im G^{R}_{\mathbf{k}'}  
\end{equation}
where both the two-particle irreducible vertex and the averaged one-electron propagator are unrestricted. The real part of the self-energy is then calculated from the Kramers-Kronig relation.\cite{Janis05b} 

This way of making the approximation with a non-local irreducible vertex compatible with the diffusion pole in the electron-hole correlation function is rather tedious and numerically demanding. Non-local parts in momenta and frequency of the self-energy are interconnected and must be determined simultaneously. There is, however, a more effective way how to assure a qualitatively correct low-energy asymptotics of function $\overline{\Phi}^{RA}$ from Eq.~\eqref{eq:EH-CF}. Instead of correcting the self-energy we can rescale the local static vertex by a positive constant so that the diffusion pole  in function $\overline{\Phi}^{RA}_E(\mathbf{q})$ is guaranteed with the local mean-field self-energy. For this purpose we introduce a positive number $0\le \phi \le 1$ with which we rescale the static part of the mean-field vertex 
\begin{equation}\label{eq:Vertex-Scaling}
\gamma^{RA} \longrightarrow \phi \gamma^{RA} 
\end{equation}
in Eq.~\eqref{eq:A-inverse}. Note that vertex $\gamma$ is not rescaled in the definition of function $S(\mathbf{q})$ in Eq.~\eqref{eq:S_function-def}, since it is consistently defined within the mean-field approximation.   The scaling parameter is determined so that the denominator vanishes in the static and homogeneous limit. That is
\begin{equation} \label{eq:DP-Stability}
1 = \phi \gamma^{RA}\left\langle \overline{G}^{A} L^{RA} \overline{G}^{R}\right\rangle \ .
\end{equation}
It is easy to rewrite this equation to a more explicit one
\begin{align} \label{eq:phi-Equation}
\phi &= \frac{\left\langle \overline{G}^{A} \overline{G}^{R}\right\rangle }{\left\langle \overline{G}^{A}L^{RA} \overline{G}^{R}\right\rangle }\nonumber \\
 &= \frac{\displaystyle \int \frac{d\mathbf{x}}{(2\pi)^{d}}\widetilde{\overline{G}}\phantom{}^{R}(\mathbf{x}) \widetilde{\overline{G}}\phantom{}^{A}(\mathbf{x})}{\displaystyle \int \frac{d\mathbf{x}}{(2\pi)^{d}} \frac{\widetilde{\overline{G}}^{R}(\mathbf{x}) \widetilde{\overline{G}}^{A}(\mathbf{x})}{1 - \phi \gamma^{RA}\overline{\chi}^{RA}\widetilde{S}^{RA}(\mathbf{x})}}
\end{align}
the solution of which  can be straightforwardly found numerically. The introduced static scaling factor effectively decreases the disorder strength and prevents the system from  undergoing  the Anderson localization  transition to the phase without macroscopic diffusion. This simplified approach to make approximations free of unphysical behavior at any disorder strength is not good enough to take into account spatial coherence of electron scatterings on random impurities beyond the weak-localization effect covered by the leading-order vertex corrections.  To reach the Anderson localization transition one has to  introduce a self-consistent renormalization of the irreducible vertex $\Lambda$.\cite{Janis05b} Nevertheless, the proposed approximation with non-local vertex corrections to the two-particle irreducible vertex does not result in unphysical behavior. 

With the diffusion pole in the electron-hole correlation function we can determine its low-energy behavior expressed in constants $A_{E}$ and $D_{E}^{0}$. They are
\begin{subequations}
\begin{multline}\label{eq:A-def}
A_{E} = -i\phi  \frac{\partial}{\partial \omega}\left[\gamma^{RA}(E+\omega,E) \left\langle \overline{G}^{A}(E) \right.\right. \\  \left.\left. 
\times L^{RA}(E + \omega,E)\overline{G}^{R}(E + \omega)\right\rangle\right]\Big|_{\omega=0} 
\end{multline}
\begin{multline}
\label{eq:D0-def}
D^{0}_{E}  = - \phi\gamma^{RA}(E,E)\\ \times  \nabla^{2}_{q}\left\langle \overline{G}^{A}(E)L^{RA}(E,E)\overline{G}^{R}(E) \right\rangle(\mathbf{q})\Big|_{q=0}
\end{multline}
\end{subequations}
where we explicitly specified the energy variables in the one and two-electron functions so that to avoid confusion. In the mean-field solution with the local irreducible vertex where the Ward identity is obeyed for all frequencies the diffusion constant is related to the static optical conductivity  at zero temperature as $\sigma = e^{2}\rho_{F}D^{0}_{F}$, where $\rho_{F}$ is the density of states at the Fermi energy.\cite{Janis03a} Since the theories with non-local irreducible vertex cannot comply with the Ward identity beyond the static limit, the diffusion constant and the conductivity will no longer obey this relation.

\section{Results}
\label{sec:Results}

\begin{figure}
  \includegraphics[scale=0.7]{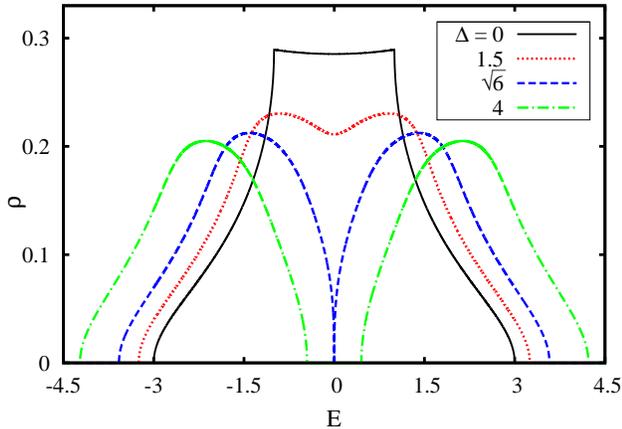}
  \caption{(Color online) Density of states of a symmetric binary alloy ($c= 1/2$) on a simple-cubic lattice for various strengths of disorder.  Split band  occurs at $\Delta = \sqrt{w}$ where $w= 6$ is the band width.    \label{fig:dos-3DBA}}\end{figure}

We use the developed approximate theory to evaluate the impact of vertex corrections on the electrical conductivity of a binary alloy on a simple-cubic lattice with a dispersion relation $\varepsilon(\mathbf{k}) = - (\cos k_{x} + \cos k_{y} + \cos k_{z})$. The site-independent distribution of the random potential is $\rho(V) = (1 - c)\delta(V - \Delta/2) + c \delta(V + \Delta/2)$.  We use the mean-field approximation (CPA) on the self-energy and the one-electron Green function. We use elliptic integrals to calculate integrals with the density of states of a simple-cubic lattice.\cite{Joyce94} Dependence of the density of states on the disorder strength is plotted in Fig.~\ref{fig:dos-3DBA}. The small imaginary part of energy  to regularize integrals along the real axis was chosen $\eta = 10^{-8}$. 

We need direct momentum integration to calculate two-particle properties in this approximation. We use the standard technique\cite{Monkhorst76} to discretize the irreducible Brillouin zone of the single-cubic lattice with $8^{3}$ to $24^{3}$ mesh points according to the precision needed.  All calculations were performed at zero temperature for a simple-cubic lattice and a symmetric ($c= 1/2$) binary alloy with a split-band limit for $\Delta_{c}= \sqrt{6}$  at which the imaginary part of the self-energy diverges.  

\begin{figure}
  \includegraphics[scale=0.75]{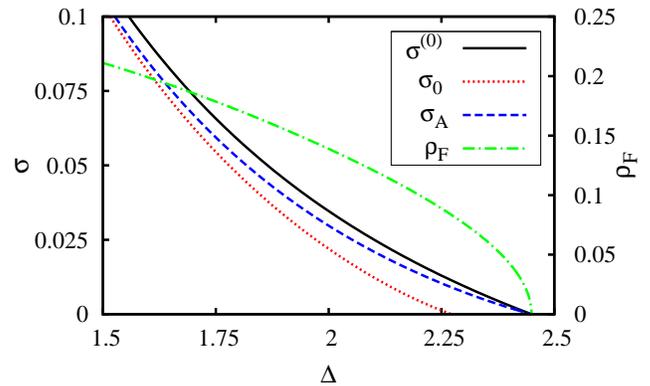}
  \caption{(Color online) Conductivities of a symmetric binary alloy on a simple-cubic lattice calculated from the Drude formula, $\sigma^{(0)}$, Drude with leading-order vertex corrections, $\sigma_{0}$, and conductivity $\sigma_{A}$ from Eq.~\eqref{eq:sigma-final} as a function of disorder strength $\Delta$. \label{fig:cond-3DBA}}\end{figure}

We first compare the results for the electrical conductivity calculated from different mean-field approaches with vertex corrections. In particular, we compare the approximate zero-temperature conductivity from Eq.~\eqref{eq:sigma-final} with that from the Drude formula, Eq.~\eqref{eq:Conductivity-Drude} and with the conductivity including vertex corrections from Eq.~\eqref{eq:Conductivity-correction} using the vertex function from the high-dimensional asymptotics, Eq.~\eqref{eq:vertex-high}. We plotted in  Fig.~\ref{fig:cond-3DBA} behavior of these conductivities  near the internal band edge (split band).  We set the electron charge $e=1$. We can see that before the split band limit is reached, the effect of disorder becomes so strong that the vertex corrections, when added linearly, overweigh the mean-field result and turn the full conductivity negative and the approximation inapplicable. The approximate non-perturbative inclusion of vertex corrections from Eq.~\eqref{eq:sigma-final} remains positive up to the band edge. Deeper within the band the two approximate formulas coincide, since the relative weight of the vertex corrections is small. 

\begin{figure}
  \includegraphics[scale=0.7]{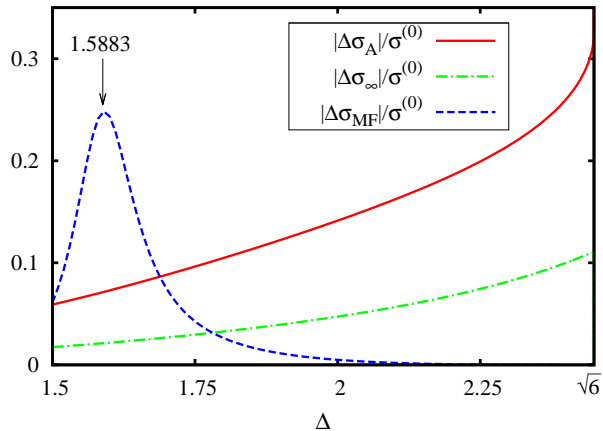}
  \caption{(Color online) Ratio of the modulus of the vertex correction and the mean-field conductivity for conductivity from Eq.~\eqref{eq:sigma-final}, $\Delta\sigma_{A}$, from Ref.~\onlinecite{Janis10}, $\Delta\sigma_{\infty}$, and from Ref.~\onlinecite{Janis01b}, $\Delta\sigma_{MF}$.  See the discussion in text.\label{fig:ratio-3DBA}}\end{figure}

The ratio of the modulus of the vertex correction and the Drude conductivity in shown Fig.~\ref{fig:ratio-3DBA}. We compared the leading-order vertex correction derived from conductivity $\sigma_{A}$ calculated from formula~\eqref{eq:sigma-final}, that is $\Delta\sigma_{A}=  \sigma_{A}- \sigma^{(0)}$. We also added the leading-order  vertex corrections in high spatial dimensions $\Delta\sigma_{\infty}$ from Ref.~\onlinecite{Janis10} and  $\sigma_{MF}$ a mean-field conductivity with the vertex corrections from Ref.~\onlinecite{Janis01b} where integration over momenta was simplified via the limit to high spatial dimensions. The approximation from Ref.~\onlinecite{Janis01b} is similar to the non-perturbative approach of this paper and was derived for the same purpose: to produce non-negative conductivity.  It completely eliminates, unlike the present one,  the divergence in the electron-hole  vertex function due to the simplification of momentum convolutions used there. We can see that the latter vertex correction shows a different trend from the other two, that is decreasing of the relative weight of vertex corrections when approaching the split-band limit. Moreover, it displays a spurious peak being a consequence of using the local irreducible CPA vertex $\lambda$  in the expansion around the mean-field solution. When rewriting the expansion in terms of the full local CPA vertex $\gamma$ this peak vanishes. This peak is a reminiscence of a singularity in the irreducible vertex $\lambda^{RR}$  for $\Delta\approx 1.5883$  that is, however, compensated in the full vertex $\gamma^{RR}$. This fact gives evidence that one has to use the full local vertex in the expansion around mean-field using the off-diagonal one-electron propagator as a small parameter.  Otherwise we have no guarantee that the singularity from $\lambda^{RR}$ is completely compensated. 

\begin{figure}
  \includegraphics[scale=0.7]{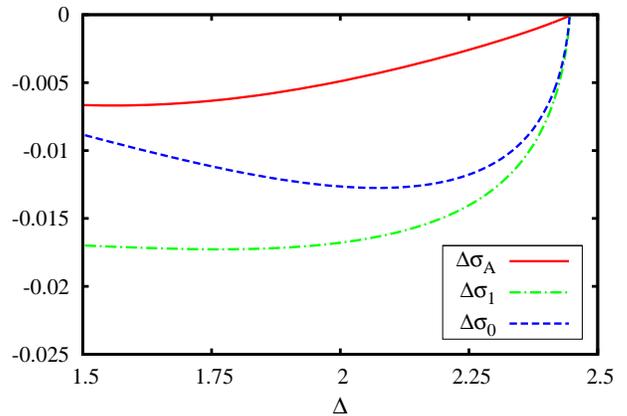}
  \caption{(Color online) Vertex corrections to the conductivity calculated from Eq.~\eqref{eq:sigma-final}, $\Delta\sigma_{A}$, first-order approximation in $\gamma$ to this  expression, $\Delta\sigma_{0}$, and the leading-order correction from the high-dimesional vertex from Eq.~\eqref{eq:vertex-high},  $\Delta\sigma_{1}$.\label{fig:vertex-3DBA}}\end{figure}

A simplification in the momentum dependence of the integral kernel $\widetilde{\overline{\Lambda}}$ from Rq.~\eqref{eq:sigma-bar-nonperturbative} is part of the non-perturbative approximation for the conductivity with vertex corrections, Eq.~\eqref{eq:sigma-final}.   We replaced a product of two off-diagonal one-electron propagators by its average over the Brillouin zone. In Fig.~\ref{fig:vertex-3DBA} we compared the vertex correction from Eq.~\eqref{eq:sigma-final} with its second order expansion in $\gamma$ (first order vanishes), $\Delta \sigma_{0}$, and the exact first-order vertex correction from Eq.~\eqref{eq:Conductivity-correction},  $\Delta \sigma_{1}$. The latter two do not differ much in the strong-disorder limit, near the split-band transition where the vertex correction is comparable with the Drude term.  For weak disorder, where the influence of the pole in function $S(\mathbf{q})$ is smaller, approximate non-perturbative and perturbative vertex corrections almost equal and differ from the exact one. The impact of the vertex correction on the total conductivity is, however, negligible there. 

\begin{figure}
  \includegraphics[scale=0.75]{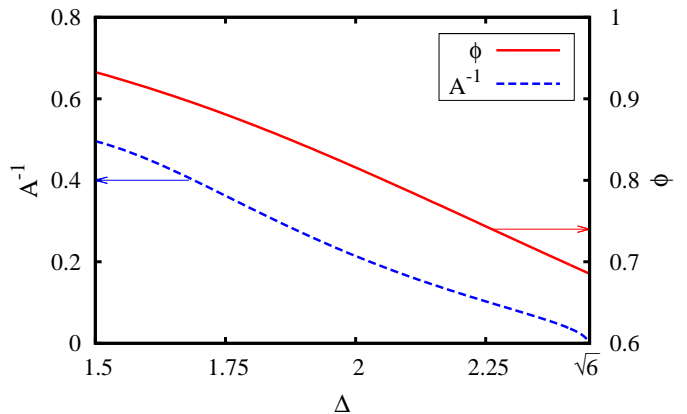}
  \caption{(Color online)  Scaling factor $\phi$  from Eq.~\eqref{eq:Vertex-Scaling} and the weight of the diffusion pole $A^{-1}$. \label{fig:A-3DBA}}\end{figure}

The primary objective of the presented construction was to rewrite and approximate the Kubo formula for the conductivity in such a way that two-particle vertex corrections to the one-electron Drude contribution never turn the total conductivity negative. Since the approximate construction uses the full two-particle vertex, there is no obstacle to use this scheme as a global approximation for one and two-particle quantities. One such a function of importance is the electron-hole correlation function $\Phi^{RA}$. It is singular in the low-energy limit if the Ward identity is obeyed on a finite frequency interval around zero.  Then also an Einstein relation couples the zero-temperature conductivity with the diffusion constant, $\sigma= e^{2}\rho_{F} D$. As we argued, validity of the full scale Ward identity cannot be reached in theories distinguishing electron-electron and electron-hole scatterings. We can only guarantee validity of the Ward identity in the static limit $\omega\to 0$. We introduced a reduction factor $\phi$ with which we secured the existence of the diffusion pole in the electron-hole correlation function without the necessity to correct the self-energy. We plotted in Fig.~\ref{fig:A-3DBA} this factor together with the weight of diffusive (extended) states $A^{-1}$. The reduction factor $\phi$ does not differ much from unity even in the strong-disorder limit. The corrections to the self-energy induced by non-locality of the irreducible electron-hole vertex that had to be included in a fully self-consistent approach are hence not of principal relevance.  As expected, the weight of the extended states vanishes at the split-band limit, $A^{-1}= 0$.  

\begin{figure}
  \includegraphics[scale=0.7]{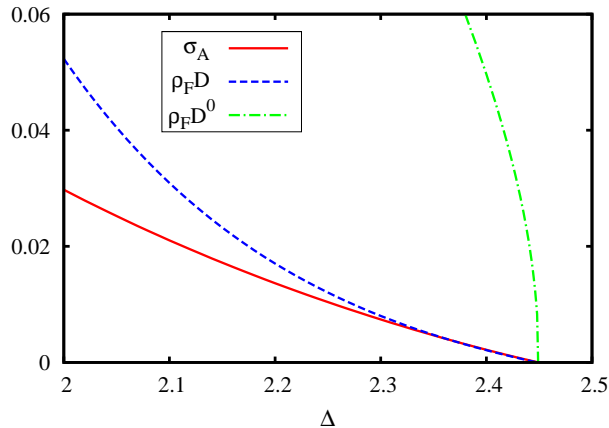}
  \caption{(Color online) Conductivity and bare ($D^{0}$) and renormalized ($D= D^{0}/A$) diffusion constants multiplied by density of states at the Fermi energy $\rho_{F}$. \label{fig:diff-3DBA}}\end{figure}

The approximate conductivity is compared with those calculated via the Einstein relation from the diffusion constants $D^{0}$ and $D= D^{0}/A$ is plotted in Fig.~\ref{fig:diff-3DBA}. We can see that in the strong-disorder limit the conductivity from the renormalized diffusion constant $D$ matches the approximated one from the Kubo formula, while the conductivity from the bare diffusion constant $D^{0}$ is quite off for all disorder strengths.  Note that $D^{0}> 0$ at the split-band limit. Conductivity $\sigma_{A}$ and $\rho_{F}D^{0}/A$ differ for intermediate disorder strengths but not (qualitatively) significantly. In the weak-disorder limit all three quantities share the same asymptotics.  

\section{Conclusions}
\label{sec:Conclusions}

We presented in this paper an approximate non-perturbative scheme for including non-local vertex corrections to the mean-field vertex from the coherent potential approximation. The construction is a partial resummation of an expansion in the off-diagonal one-electron propagator from CPA that includes the leading-order non-local correction to the irreducible electron-hole irreducible vertex. After a simplification in the momentum dependence of the integral kernel of the Bethe-Salpeter equation with multiple scatterings we derived an explicit representation for the electrical conductivity, Eq.~\eqref{eq:sigma-final}, that remains non-negative irrespectively of how strong the disorder is.  We explicitly demonstrated that this expression produces reliable results for the electrical conductivity, in particular in the strong-disorder regime, and it qualitatively correctly mimics the behavior of the averaged conductivity in intermediate and weak-disorder regimes. An important aspect of the derived approach is that we made the approximation for the non-local part of the two-particle vertex compatible with the local mean-field one-electron self-energy. It means that it is not necessary to correct the self-energy by adding non-local terms in a cumbersome way to assess the averaged conductivity with vertex corrections. The only modification to be done in the electron-hole vertex is to rescale the local mean-field vertex $\gamma$ by a factor so that to restore the diffusion pole in the electron-hole correlation function. This scaling effectively reduces the disorder effect in the regions where the vertex corrections to the conductivity overweigh the Drude term and one needs to apply a non-perturbative approach to correct this deficiency. Our construction does exactly this with a minimum effort and no changes in the mean-field one-electron self-energy.  It can hence be used to extend the standard mean-field  one-electron approximation by including leading two-particle vertex corrections to the electrical conductivity without breaking consistency and the fundamental physical laws. The approximate construction is enough simple so that it may find application in the calculation of transport properties of real disordered materials beyond the model level presented here.  

\section*{Acknowledgement} %
 VP acknowledges partial support by  grant SVV 265 301 of Charles University in Prague.

\section*{References}
%\medskip

\end{document}